\documentstyle[12pt]{article}
\renewcommand{\deg}{^\circ}
\newcommand{\comment}[1]{}
\newcommand{\htadthree}{H^3\wsub{tad}}
\newcommand{\rip}[1]{\left| {#1} \right\rangle}

\newcommand{\trio}[3]{{\pil{ #1 \abs{ #2 } #3}}}

\newcommand{\pil}[1]{{\left\langle #1 \right\rangle}}

\newcommand{\pip}[1]{{\left( #1 \right)}}

\newcommand{\abs}[1]{{\left| #1 \right|}}
\newcommand{\w}[1]{{\mbox {#1}}}
\newcommand{\wsub}[1]{_{{\mbox {\scriptsize {#1}}}}}

\newcommand{\ov}[1]{{\overline {#1}}}
\newcommand{\eqnb}{\begin{equation}}
\newcommand{\eqne}{\end{equation}}
\newcommand{\fpi}{{f_{\pi}}}

\newcommand{\reo}{{\rho\deg\w{--}\omega}}
\newcommand{\questiona}{{\nu}}
\newcommand{\hem}{{H\wsub{em}}}
\newcommand{\ho}{{\trio{\rho\deg}{\hem}{\omega}}}
\newcommand{\dio}{{\Delta I = 1}}
\newcommand{\hjj}{{H\wsub{JJ}}}

\newcommand{\mks}[1]{m_{K^{*{#1}}}}
\title{ Resurrection of Charge Symmetry Violation for $\ho \sim
 	-5000 ~\w{MeV}^2$}
\author {M.\ D.\ Scadron $^{\dag}$\\  
	School of Physics, University of Melbourne, Parkville,\\ 
	Victoria 3052, Australia}
\date{}
\begin{document}
\setlength{\textheight}{120 ex}
\addtolength{\baselineskip}{1 ex}
\addtolength{\parskip}{1 ex}
\setlength{\topmargin}{-7 ex}
\maketitle
\vfill
\begin{center}{Abstract}\end{center}
We review the theoretical predictions for $\reo$ mixing and conclude in many ways that
$\ho \sim -5000 ~\w{MeV}^2$, more in line with earlier analyses
of charge symmetry violation and with present data.
\vfill
\vfill
\noindent
\newline
$^{\dag}$Permanent Address:  Physics Department, University of Arizona,  
Tucson, AZ 85721, USA. 
\newpage
As of the reviews of 1990 and earlier [1], charge symmetry violation
(CSV) in nuclear physics was thought to be reasonably well understood.  
This was in part due to earlier rho-omega ($\reo$) mixing analyses of 
Glashow [2], Renard [3] and Coon et al [4], culminating in the Coon and 
Barrett conclusion [5] in 1987 that the most accurate particle data 
group measurements lead to the $\dio ~~\reo$ electromagnetic 
hamiltonian density ($\hem$) transition  
$$
  \ho = - (4520 \pm 600) ~\w{MeV}^2~~.
\eqno(1)
$$
Only recently Coon and Scadron [6] have shown on the basis of
SU(3) and SU(6) symmetry, that the theoretical versions of both
nonstrange (NS) $\omega$--$\rho{^0}$ and $\eta_{NS}$--$\pi^0 ~~ \dio$ 
transitions have the {\em universal} value
$$
  \ho \approx \trio{\pi{^0}}{\hem}{\eta_{NS}} \sim
  -5000~\w{MeV}^2~,
\eqno(2)
$$
for similar mass shells 
$m_\omega \approx 783$ MeV, $m_{\rho{^0}} \approx 769$ 
MeV, $m_{\eta_{NS}} \sim \frac{1}{2} (m_{\eta'} + m_\eta) \sim
753$ MeV~.

This universal CSV scale in (2) is also compatible [6] with the 
Coleman-Glashow [7] and Dashen theorem [8] versions of all 13 
electromagnetic (em) ground state pseudoscalar (P), vector (V), 
octet baryon (B) and decuplet baryon (D) SU(2) mass splittings [9]. 
Further contact with CSV data follows from constructing [1]
class III and class IV $NN$ potentials from the off-shell spacelike
$q^2$ CSV $\dio$ exchange graph of Fig.\ 1.  These Fourier-transformed 
potentials are proportional to $e^{-m_V r}$ or $e^{-m_V r}/r$, 
where $m_V= (m_\rho + m_\omega)/2$.   This class III potential 
is compatible with $NN$ scattering and bound state (Nolen-Schiffer anomaly) 
data in mirror nuclear systems [1,5,10], with Coulomb displacement energies of 
isobaric analog states [11], and with isospin-mixing matrix elements 
relevant to the isospin-forbidden beta decays [12].  The class IV CSV 
potential [13] is compatible with precise measurements of the elastic 
scattering of polarized neutrons off of polarized protons [14].

Not withstanding the above rather complete picture of CSV, beginning 
with ref.\ [15] in 1992, a new approach to CSV in nuclear physics has 
begun to emerge [16--19].  This approach questions [15--19] the validity 
of both the empirical and theoretical values of the $\reo$ mixing 
transitions in (1) and (2) while claiming instead that there is further 
off-shell $q^2$ dependence of $\reo$ mixing beyond the usual 
Fourier-transformed $NN$ potential following from Fig.\ 1.

	In this letter we briefly summarize why these new CSV questions since 1992 [15--19] are misdirected and do not negate the (successful) CSV $\reo$ mixing picture given in our first two paragraphs.

	To begin, we remind the reader that the $\dio$ Coleman-Glashow (CG) em hamiltonian density in (1) or (2) has (photon) current-current (JJ) and contact (tadpole) parts
$$
  \hem = \hjj + {\htadthree}~~.
\eqno(3)
$$
While both matrix elements of $\hjj$ and of ${\htadthree}$ are necessary to fit all 13 [6] P, V, B, D em mass differences of (3), only the contact (tadpole) part in (3) is needed for the $\reo$ mixing transition, leading to
$$
  \ho = \trio{\rho{^0}}{{\htadthree}}{\omega} =
  \Delta \mks{}^2 - \Delta m_{\rho}^2 \approx -4666 ~\w{MeV}^2~.
\eqno(4)
$$
Here we have assumed SU(3) symmetry for an $\omega$ as pure 
nonstrange $(\ov{u} u + \ov{d}d) / \sqrt{2}$ and used the 1996 PDG [9] 
$K^*$ and $\rho$ masses, $\mks{+} \approx 891.6$ MeV, $\mks{0} \approx 
896.1$ MeV so that $\Delta \mks{}^2 = \mks{+}^2 - \mks{0}^2 \approx -8045 
~\w{MeV}^2$, along
with $m_{\rho^+} \approx 766.9$ MeV, $m_{\rho{^0}} \approx 769.1$ MeV, so 
that $\Delta m_\rho^2 = m_{\rho^+}^2 - m_{\rho{^0}}^2 \approx -3379 
~\w{MeV}^2$. This results in the $\reo$ mixing scale given in eq.\ (4).

	The above CG contact (tadpole) $\reo$ mixing prediction (4) must hold by virtue of the Dashen observation [8] $\trio{\pi{^0}}{\hjj}{\eta} = 0$ extended to $\trio{\rho{^0}}{\hjj}{\omega} = 0$ [6].  But it is important to demonstrate
that the contact-tadpole graph of Fig.\ 2 actually recovers the
$\reo$ mixing scales of eqs.\ (1), (2), (4).  Here the $\Delta I = 1 ~\delta{^0}$ pole in Fig. 2 is now called [9] the $a_0^0$ (984).  Reference [6] shows
in fact that
$$
  \trio{\rho{^0}}{{\htadthree}}{\omega} \approx - f_\delta ~ 2g_{\delta\omega\rho}
 \approx - f_\delta m_\delta^2 / \fpi \approx -4400~\w{MeV}^2~,
\eqno(5)
$$
where the ``decay constant'' $f_\delta$ is 0.42 MeV as found from
the SU(2) to SU(3) symmetry breaking scale of 2\%
where $\fpi \approx 93 ~\w{MeV}$.  We believe that the close agreement between the $\reo$ mixing scales in (1), (2), (4) and (5) are significant---they justify the CG tadpole picture (as do the successful 13 P, V, B, D em mass splittings).

	By way of contrast, ref.\ [15] ignores this CG contact tadpole picture and instead studies the (non-contact) $\reo$ GHT [15] quark loop of
Fig.\ 3 (their Fig.\ 1), where $\delta m = m_u - m_d$ corresponds to the
$u$--$d$ current quark mass difference.  While it is true [6] that $\delta
m$ appears in the CG tadpole hamiltonian ${\htadthree} = \delta m (\ov{u}u
- \ov{d}d) / 2$, the GHT
quark loop of Fig.\ 3 is {\em not} a CG contact-tadpole of Figs.\ 1 or 2.  
Moreover the GHT quark loop picture does not recover 
the 13 P, V, B, D em mass differences nor the $\ho$ $\Delta I = 1$ 
scales of eqs.\ (1), (2), (4), (5) above.

	In spite of these (serious) deficiencies, refs.\ [16--19] continue 
to study the properties of Fig.\ 3 and learn that nonstrange quark 
loops have a large off-shell $q^2$ dependence when coupled to 
external $\rho{^0}$ and $\omega$ states.  Whether this is true or 
not is irrelevant because such GHT nonstrange $u$ and $d$ quark loops 
cannot generate the needed CG contact-tadpole graphs of 
Figs.\ 1 or 2, which explain CSV due to the $\dio$ $\reo$ contact transition.

	As noted in the second reference in [4], the ``boosted'' form of the contact $\ho$ transition is

$$
  \varepsilon^{\mu^*} {{\trio{{\rho_\mu}\deg}{\hem}{\omega_\nu}}} 
\varepsilon^{\questiona} =
  \varepsilon^{\mu^*} \pip{g_{\mu \nu} - \frac{q_\mu q_{\questiona}}
	{t}} \varepsilon^{\questiona} \ho~,
\eqno(6)
$$
where $t = q^2$ is the invariant squared momentum of the vector mesons.
Because of the conserved isovector $\rho{^0}$ current, the second
term on the right-hand side of (6) vanishes and there is no $q^2$
variation of this contact $\reo$ mixing transition (6) since the $\omega (783)$ and 
$\rho (770)$ are essentially on the same mass shells.

Even if this GHT quark loop were the origin of $\reo$ mixing (which
is not physically possible as explained above), a recent analysis [20] using Borel and finite energy QCD sum rules shows that the inclusion of finite
mesonic widths requires the $\reo$ mixing matrix element in the
space-like region to have the same sign and similar magnitude
as its on-shell value, eqs.\ (1), (2), (4), (5) above.  Moreover,
ref.\ [20] demonstrates that the ``node theorem'' of ref.\ [17] for the
mixed $\rho{^0}$ and $\omega$ propagator is spoiled.

	This misuse of the ``mixed $\rho{^0}$ and $\omega$ propagator''
culminated in the recent analysis [19] of $e^+ e^- \to \pi^+ \pi^-$ in
the $\reo$ interference region, claiming that data from the 
$e^+ e^- \to \pi^+ \pi^-$ interference region cannot be used to fix the
value of $\reo$ mixing in a model-independent way.
This conclusion is in contrast with eqs.\ (1), (2), (4), (5) above and
follows from a misinterpretation of the contact-tadpole analysis of 
Renard [3] calling there $\delta m$ [21], but meaning the CG
contact-tadpole $\ho$~.

	To make our point in more detail, we follow refs.\ [4] and Fig.\ 4a to compute $\ho$ using the 1996 PDG rates [9] $\Gamma (\rho{^0} \to 2\pi) = 150.7 \pm 1.2 ~\w{MeV}$ and $\Gamma (\omega \to 2\pi) = 0.186 \pm 0.025 ~\w{MeV}$:  	
$$
  \Gamma (\omega \to 2 \pi) \approx \left| \frac{\ho}{im_\rho \Gamma_\rho} \right|^2 \Gamma (\rho{^0} \to 2\pi)~,
\eqno(7a)
$$
$$
  \ho \approx - (4069 \pm 277) ~\w{MeV}^2~.
\eqno(7b)
$$
The sign of (7b) is uniquely determined from the $\rho{^0}$ and 
$\omega$ interference phase in $e^+ e^- \to \pi^+ \pi^-$ of $\phi
\sim 100{^\circ}$ [4], where this phase was also found by Renard [3].  
A judicious choice of $\omega \to 2\pi$ data increases this scale in (7b) 
to the value [5] in (1).  This $\omega$--$\rho$ transition in (1) or 
in (7b) is a Coleman-Glashow [7] shaded contact (tadpole) 
interaction of Figs. 1,2,4; it is {\em not} a (non-contact) scale arising 
from a $\reo$ mixed quark propagator or from the quark bubble of Fig.\ 3.

	The notion of quantum-mechanical mixing applies to strong interaction $\phi$--$\omega$ mixing with states
$$
  \rip{\omega} = \cos{\phi_V} \rip{\omega_{NS}} - \sin{\phi_V}
	\rip{\omega_S}
\eqno(8a)
$$
$$
  \rip{\phi} = \sin{\phi_V} \rip{\omega_{NS}} + \cos \phi_V \rip{\omega_S}
\eqno(8b)
$$
relative to the nonstrange $\rip{\omega_{NS}} = (\rip{\ov{u}u} +
\rip{\ov{d}d}) / \sqrt{2}$ and strange $\rip{\omega_S} = \rip{\ov{s}
s}$ quark basis [22].  This mixing angle $\phi_V$ is very small [23] and can be obtained from data using [9] $\Gamma (\phi \to \pi \gamma) =
5.80 \pm 0.58$ keV and $\Gamma (\omega \to \pi \gamma) = 717 \pm
43$ keV:  
$$
  \frac{\Gamma (\phi \to \pi \gamma)}{\Gamma (\omega \to \pi \gamma)} =
  \frac{p_\phi^3}{p_\omega^3} \tan^2 \phi_V =
  0.00809 \pm 0.00094~,
\eqno(9)
$$
requiring $\phi_V \approx 3.4{^0}$ for $p_\phi = 501$ MeV, $p_\omega = 379$ MeV.  There is no point in finding a rediagonalized $\reo$ (or $\pi{^0}$--$\eta_{NS}$) mixing angle, but it is small anyway [21].

	To demonstrate that this $\phi$--$\omega$ mixing angle analysis
of (8) and (9) can also be used to compute the $\reo$ transition of 
interest (here we prefer not to call this transition $\reo$ ``mixing''),
we see from eqs.\ (8a) and (7) that
$$
  \trio{\rho{^0}}{\hem}{\omega_{NS}} = \frac{1}{\cos \phi_V}
  \ho \approx - 4076~\w{MeV}^2~,
\eqno(10)
$$
near (7b) because $\omega$ is 97\% nonstrange.  Likewise the
$\phi \to 2\pi$ rate [9] found from the total $\phi$ rate
$4.43 \pm 0.05$ MeV and branching ratio $(8 \pm 4) \cdot 10^{-5}$
gives the $\rho{^0}$--$\phi$ contact em transition of Fig.\ 4b as
$$
  \Gamma (\phi \to 2\pi) \approx
  \left| \frac{{\trio {\rho\deg}{\hem}{\phi}}}{im_\rho \Gamma_\rho}
\right|^2 \Gamma(\rho \to 2\pi)
\eqno(11a)
$$
or $\langle \rho\deg | \hem | \phi \rangle \approx -178 ~\w{MeV}^2$.  Although this latter $\dio$ em 
transition is $\sim 30$ times smaller than (1) or (10), the small observed
mixing angle
$\phi_V \approx 3.4{^\circ}$ enhances this scale to
$$
  \trio{\rho{^0}}{\hem}{\omega_{NS}} = \frac{1}{\sin \phi_V}
  \trio{\rho{^0}}{\hem}{\phi} \approx -3020 \pm 710 ~\w{MeV}^2~.
\eqno(11b)
$$
The latter large error suggests (11b) is compatible with the
$\reo$ $\dio$ scale in (10) or in (1), (2), (4), (5).  
If we select
the Vasserman et al $\phi \to 2\pi$ data [9] then (11b) increases
to 
$$
  \trio{\rho{^0}}{\hem}{\omega_{NS}} \approx -4700 ~\w{MeV}^2,
\eqno(11c)
$$
more in line with (1), (2), (4), (5).

This $\phi$--$\omega$ mixing angle digression in eqs.\ (8)--(11)
consistently using a $\dio$ {\em contact} (tadpole)  along with $\reo$ and
$\rho{^0}$--$\phi$ transitions is in direct conflict with ref.\ [19]
(and also [15]--[18]).  They deal with (unphysical) non-contact
quark loop graphs and a mixed $\reo$ off-diagonal propagator  while
assuming (with no support from data) that the pure isoscalar omega has
an isospin violating coupling to two pions and 
indirectly suggest that the result (11) could not possibly hold.

In this letter we have briefly reviewed {\em four} determinations of 
the charge symmetry violating (CSV) scale $\ho \sim -5000 ~\w{MeV}^2$ in
equs. (1), (2), (5), (7) while giving {\em two} new derivations of 
this CSV scale in equs. (4) and (11).

These {\em six} above determinations of $\ho$ all 
assume that nonphotonic $\Delta I = 1$ transitions are always of the 
contact-tadpole $u_3 =\bar{q}\lambda{_3}q$ form (proportional to
${\htadthree}$ in this letter) as originally 
proposed by Coleman and 
Glashow [CG] and supported by Weinberg [7]. This $u_3$ contact term is 
represented by the shaded $\Delta I = 1$ interactions in our Figs.1,2,4.
Support for this CG scheme is given by the universal electromagnetic
SU(2) mass splittings of all (13) ground state P,V,B,D hadrons as 
summarized in the appendix of ref. [6.]    
However, this successful analysis is not related to the quark-loop mixed
$\reo$ propagator formalism recently developed in [15--19].
	
	In conclusion, we believe that charge symmetry violation in nuclear
physics is well understood [1--6].  It is based on the particle 
physics $\dio$ $\reo$ contact-tadpole transition [6--8], which is also 
well understood.

{\bf Acknowledgements}: 
\newline
The author appreciates the hospitality of the University of
Melbourne.
\section*{Figure Captions}
\begin{description}
\item{Fig.\ 1} Nucleon-nucleon CSV vector meson exchange graph.
\item{Fig.\ 2} Delta meson tadpole dominance of the CSV $\dio ~
  \ho$ contact transition.
\item{Fig.\ 3} Nonstrange quark loops for $\reo$ mixing.
\item{Fig.\ 4} $\omega$--$\rho{^0}$ $\dio$ contact graph dominating
  $\omega \to 2\pi$ (a);  $\phi$--$\rho{^0}$ $\dio$ contact graph
  dominating $\phi \to 2 \pi$ (b).
\end{description}
\newpage
\section*{References}
\begin{description}
\item{1.}
See reviews of E.\ M.\ Henley and G.\ A.\ Miller, in
{\em Mesons and Nuclei}, edited by M.\ Rho and D.\ Wilkinson
(North-Holland, Amsterdam, 1979); G.\ A.\ Miller, B.\ M.\ K.\ 
Nefkens and I.\ \u{S}laus, Phys. Rept. {\bf 194} (1990) 1.

\item{2.}  
S.\ L.\ Glashow, Phys.\ Rev.\ Lett.\ {\bf 7} (1961) 469.

\item{3.} 
F.\ M.\ Renard, Springer Tracts in Modern Physics
{\bf 63} (1972) 98.

\item{4.} 
P.\ C.\ McNamee, M.\ D.\ Scadron and S.\ A.\ Coon, Nucl.\ Phys.\ A {\bf 249} 
(1975) 483; {\bf 287} (1977) 38; S.\ A. \ Coon and M.\ D.\ Scadron, Phys.
Rev. C {\bf 26} (1982) 562.

\item{5.} 
S.\ A.\ Coon and R.\ C.\ Barrett, Phys.\ Rev.\ C {\bf 36}
(1987) 2189.

\item{6.} 
S.\ A.\ Coon and M.\ D.\ Scadron, Phys.\ Rev.\ C {\bf 51}
(1995) 2923.

\item{7.} 
S.\ Coleman and S.\ L.\ Glashow, Phys.\ Rev.\ Lett.{\bf 6} (1961) 423; 
Phys.\ Rev. {\bf 134} (1964) B671; also noted by S.\ Weinberg, Phys.\ 
Rev.\ D {\bf 11} (1975) 3583. 

\item{8.}
R.\ Dashen, Phys.\ Rev.\ {\bf 183} (1969) 1245.  The 
consequent ``Dashen Theorem'' reads ``$\Delta m_K^2 = \Delta m_\pi^2$ 
to order $e^2$, neglecting strong interaction violations of SU(3)
$\times$ SU(3)''.  These latter ``violations'' are in fact the 
leading Coleman-Glashow tadpole terms in ref.\ [7], which change
the Dashen prediction $\Delta m_K^2 = \Delta m_\pi^2$ much nearer to the 
empirical value $\Delta m_K^2 \approx -3 \Delta m_\pi^2$ [6].

\item{9.}
Particle Data Group, R.\ M.\ Barnett et al., Phys.\ Rev.\ D
{\bf 54} (1996) 1, Part I.

\item{10.}
Y.\ Wu, S.\ Ishikawa and T.\ Sasakawa, Phys.\ Rev.\ Lett.\
{\bf 64} (1990) 1875;
T.\ Suzuki, H.\ Sagawa and A.\ Arima, Nucl.\ Phys.\ A {\bf 536}
(1992) 141. Also see P.G. Blunden and M.J. Iqbal, Phys. 
Lett.B{\bf 198} (1987) 14.

\item{11.}
T.\ Suzuki, H.\ Sagawa and N.\ Van Gai, Phys.\ Rev.\ C
{\bf 47} (1993) R1360.

\item{12.}
S.\ Nakamura, K.\ Muto and T.\ Oda, Phys.\ Lett.\
B {\bf 297} (1992) 14; {\bf 311} (1993) 15.

\item{13.}
A.\ G.\ Williams, A.\ W.\ Thomas and G.\ A.\ Miller,
Phys.\ Rev.\ C {\bf 36} (1987) 1956.

\item{14.} 
L.\ D.\ Knutson et al, Phys.\ Rev.\ Lett. {\bf 66}
(1991) 1410; S.\ E.\ Vigdor et al, Phys.\ Rev.\ C {\bf 46}
(1992) 410.  See also R.\ Abegg et al, Phys.\ Rev.\ Lett. {\bf 56}
(1986) 2571;  Phys.\ Rev.\ D {\bf 39} (1989) 2464; Phys.\ Rev.\ 
Lett.{\bf 75} (1995) 1711.

\item{15.}
T.\ Goldman, J.\ A.\ Henderson and A.\ W.\ Thomas,
Few Body Systems {\bf 12} (1992) 123 (GHT).

\item{16.}
G.\ Krein, A.\ W.\ Thomas and A.\ G.\ Williams, Phys.\
Lett.\ B {\bf 317} (1993) 293; J.\ Piekarewiez and A.\ G.\ Williams,
Phys.\ Rev.\ C {\bf 47} (1993) R2461; T.\ Hatsuda, E.\ M.\ Henley, T.\ Meissner and G.\ Krein, ibid.\ C {\bf 49} (1994) 452; K.\ L.\ Mitchell, P.\ C.\
Tandy, C.\ D.\ Roberts and R.\ T.\ Cahill, Phys.\ Lett.\ B
{\bf 335} (1994) 282.

\item{17.}
H.\ B.\ O'Connell, B.\ C.\ Pearce, A.\ W.\ Thomas and A.\ 
C.\ Williams, Phys.\ Lett. B {\bf 336} (1994) 1;
ibid.\ B {\bf 354} (1995) 14.

\item {18.}
R.\ Urech, Phys.\ Lett.\ B {\bf 355} (1995) 308; 
A.\ N.\ Mitra and K.\ C.\ Yang, Phys.\ Rev.\ C {\bf 51} (1995)
3404; K.\ Maltman, Phys.\ Lett.\ B {\bf 362} (1995) 11.

\item {19.}
K.\ Maltman, H.\ B.\ O'Connell and A.\ G.\ Williams,
Phys.\ Lett.\ B {\bf 376} (1996) 19.

\item{20.}
M.\ J.\ Iqbal, X.\ M.\ Jin and D.\ B.\ Leinweber, Phys.\ Lett.\ B{\bf 
386} (1996) 55.

\item{21.}
M.\  Gourdin, Unitary Symmetries and their Application
to High Energy Physics (North-Holland, Amsterdam, 1967),
pp.\ 91--101.

\item{22.}
For mixing in the $NS$--$S$ (quark) basis see H.\ F.\ 
Jones and M.\ D.\ Scadron, Nucl.\ Phys.\ B {\bf 155} (1979) 409;
M.\ D.\ Scadron, Phys.\ Rev.\ D {\bf 29} (1984) 2076; A.\ Bramon
and M.\ D.\ Scadron, Phys.\ Lett.\ B {\bf 234} (1990) 346.

\item{23.}
To transform from the $NS$--$S$ basis (angle $\phi$)of refs.\ [22] to the
singlet-octet basis (angle $\theta$), use $\theta = \phi - \arctan
\sqrt{2}$, the latter
angle being $54.7\deg$~.  

\end{description}
\end{document}